\documentclass{article}

\input epsf.tex

\newcommand{\be}{\begin{equation}}
\newcommand{\ee}{\end{equation}}
\newcommand{\ba}{\begin{eqnarray}}
\newcommand{\ea}{\end{eqnarray}}
\newcommand{\ban}{\begin{eqnarray*}}
\newcommand{\ean}{\end{eqnarray*}}

\newcommand{\ket}[1]{\mbox{$ | #1 \rangle $}}
\newcommand{\bra}[1]{\mbox{$ \langle #1 | $}}

\newcommand{\demi}{\frac{1}{2}}
\newcommand{\compl}{\begin{picture}(8,8)\put(0,0){C}\put(3,0.3){\line(0,1){7}}\end{picture}}

\newcommand{\one}{\leavevmode\hbox{\small1\normalsize\kern-.33em1}}

\newcommand{\Tr}{\mbox{Tr}}

\newcommand{\moy}[1]{\langle #1 \rangle}

\def\H{{\cal H}}
\def\proj#1{\ket{#1}\!\bra{#1}}

\begin{document}

\title{Bell's inequalities detect efficient entanglement}

\author{Antonio Ac\'{\i}n$^1$, Nicolas Gisin$^2$, Lluis Masanes$^3$, Valerio
Scarani$^2$\\ $^1$ Institut de Ci\`encies Fot\`oniques, Barcelona,
Spain.\\ $^2$ Group of Applied Physics, University of Geneva,
Switzerland.\\ $^3$ Dept ECM, University of Barcelona, Spain.}


\maketitle

\begin{abstract}
We review the status of Bell's inequalities in quantum
information, stressing mainly the links with quantum key
distribution and distillation of entanglement. We also prove
that for all the eavesdropping attacks using one qubit, and
for a family of attacks of two qubits, acting on half of a
maximally entangled state of two qubits, the violation of
a Bell inequality implies the possibility of
an efficient secret-key extraction.
\end{abstract}


\section{Introduction}

Quantum correlations were noticed to be astonishing by
Einstein-Podolski-Rosen \cite{EPR} and by Schr\"odinger
\cite{Schr} back in 1935. In particular, the EPR paper stressed
that the predicted correlations could not be explained by exchange
of a signal, since the entangled particles could be at an
arbitrary distance from one another. If signal exchange is
excluded, in the classical world we know only another mechanism to
establish correlations: common preparation at the source. This
second possibility was ruled out by John Bell in 1964 \cite{Bell}:
the predicted quantum correlations violate a condition (``Bell's
inequality", BI) that should hold if the correlations were
established at the preparation. All the experiments performed
since the Aspect experiment \cite{AGR} in 1982 confirm quantum
physics.

Nowadays, although one should not forget the detection and
locality loophole until their joint experimental test \cite{loop},
for most physicists the debate on quantum correlations is closed:
{\em entanglement does exist}, and moreover it has been recognized
as a {\em resource} needed to perform tasks that would be
classically impossible \cite{QIPIntro98}. While nobody doubts that
the interpretational content of the BI should shape any
physicist's view of the world, it is not clear whether BI can be
of interest for quantum information processing. We have
investigated this question, since we believe that deep concepts
and clever applications should not become two separate domains.

This paper contains two separate sections: in section
\ref{secrev}, we review the status of the Bell's inequalities in
quantum information; in section \ref{secqub}, we present a
generalization of the link \cite{fuchs} between violation of
Bell's inequalities and security of the quantum key distribution
with qubits.

\section{The status of Bell's inequalities in quantum information}
\label{secrev}

\subsection{BI and quantum cryptography}

The goal of quantum cryptography (quantum key distribution, QKD)
is to provide Alice and Bob with a secret key. An important result
of classical cryptography says that if \ba
I(A:B)&>&\min[I(A:E),I(B:E)]\,\label{ck}\ea where $I(X:Y)$ is the
mutual information between $X$ and $Y$, then a secret key can be
extracted from the classical data (obtained by measuring the
quantum systems) by efficient protocols using only one-way
communication \cite{CK}. If that condition does not hold, in some
cases a secret key can still be extracted, but all the known
protocols are very inefficient.

Consider QKD with entangled particles \cite{Ekert}, and let us
start with the standard setting with two partners. Alice prepares
the maximally entangled state, keeps one particle and sends the
other one to Bob. In the absence of any spy on the line, whenever
Alice and Bob measure in the same basis they obtain perfectly
correlated random results. If the eavesdropper Eve has her own
particles interact with the particle flying to Bob, the quantum
state $\ket{\Psi_{ABE}}$ becomes shared among the three actors,
and the quantum information shared by Alice and Bob is given by
the mixed state $\rho_{AB}$ obtained by tracing out Eve's system.
The connection with BI is as follows: in all the studied
protocols, considering Eve's optimal {\em individual} attack, if
$\rho_{AB}$ violates a Bell's inequality, then (\ref{ck}) holds.

This link was first noticed for the four-state protocol with
qubits \cite{fuchs}, in which case actually the violation of the
CHSH inequality is also a necessary condition for (\ref{ck}) to
hold. In section \ref{secqub} we generalize this result. Protocols
using higher-dimensional systems and/or more bases have also been
studied, and in all these cases the condition seems only to be
sufficient \cite{qic}.

A different extension has also been studied: the extension to
protocols involving more than two partners. In such ``quantum
secret-sharing" protocols, Alice distributes random bits to $N$
Bobs, that must cooperate in order to retrieve the key. For
protocols using qubits and two conjugated bases, it was also found
\cite{bell2} that a condition similar to (\ref{ck}) holds if and
only if the Mermin-Klyshko inequalities are violated.

\subsection{BI and distillation of entanglement}

Distillation of entanglement is a fundamental quantum information
process. The entanglement of a quantum state $\rho$ is distillable
if, out of many copies of it, one can extract maximally entangled
states (two-qubit singlets) using only local operations and
classical communication (LOCC). Operationally, this means the
following: if a source $S$ produces a state which is weakly
entangled but distillable, then one can build a new source $S\,'$,
that is less efficient but produces strongly entangled states, by
simply appending local devices to the ports of $S$ and allowing
the partners to communicate. In other words, if we have $S$, then
we can build $S\,'$ and run any quantum information protocol like
teleportation. The notion of distillability is not trivial
because, in all quantum composed systems but
$\compl^{\,2}\otimes\compl^{\,2}$ and
$\compl^{\,2}\otimes\compl^{\,3}$, there exist so-called
bound-entangled states, that are entangled but not distillable.

We studied the link with BI in quantum systems composed of $N>2$
qubits. In such a case, when the system is composed of more than
two sub-systems, the notion of distillability is not even
univoque. The strongest requirement is ``full distillability": any
two partners can distill a singlet by LOCC. The weakest
requirement is ``bipartite distillability": the $N$ partners split
into two groups of $n_A$ and $n_B=N-n_A$ partners, and the state
is distillable with respect to this partition $n_A/n_B$. Within
each group, the most general transformations are allowed; but only
classical communication is allowed between one group and the
other.

We have demonstrated \cite{main} a quantitative link between this
hierarchy or {\em degree of distillability} and the amount of
violation of the WWZB inequalities \cite{wwzb}, that are the
linear correlation inequalities with two settings per site. If a
$N$-qubit state violates a WWZB inequality there is some
distillable entanglement in the state; moreover, the amount of the
violation is associated to the degree of distillability. In
particular, a violation close to the maximal value, namely
$\moy{B_N}\in ]2^{(N-2)/2},2^{(N-1)/2}]$, guarantees full
distillability of the state. A similar result holds for the Uffink
inequality \cite{Uffink}.

\subsection{BI and communication complexity}

A ``communication complexity" problem is the problem of computing a
function whose inputs are distributed among several partners, who
can exchange only a limited amount of information. In the quantum
version of such protocols, some of the input information is
replaced by quantum information, and the partners can share an
entangled state.

It has been shown in Ref. \cite{complex} that for every Bell's
inequality and for a broad class of protocols, there always exists
a multi-partite communication complexity problem, for which the
protocol assisted by states which violate the inequality is more
efficient than any classical protocol. Moreover, for that
advantage, the violation of the BI is a necessary and sufficient
criterion.

\section{CHSH and quantum key distribution with qubits}
\label{secqub}

In this section we will show the link between Bell violation
and the security of QKD protocols for all one-qubit
eavesdropping attacks and a family of two-qubit attacks.

Consider the situation in which Alice locally prepares a maximally
entangled state of two qubits, $\ket{\Phi^+}=(\ket{00}+\ket{11})
/\sqrt 2$, and sends one of the qubits to Bob by an insecure
quantum channel. This qubit is intercepted by Eve, who performs
the following attack: (i) she adds a {\em one-qubit} ancillary
system in the state $\ket{E}$ and performs a unitary operation,
$U_{BE}$, over the two qubits and (ii) forwards one of the output
qubits to Bob. Giving Eve just one qubit may appear as an
exceedingly strong restriction; however, it is known that there
exist a one-qubit attack such that $I(A:E)$ reaches the value of
the optimal individual eavesdropping on the BB84 protocol
\cite{NG,bell2}. We shall discuss below the role of $I(B:E)$.

After this attack, Alice, Bob and Eve share a three-qubit pure
state, $\ket{\Psi_{ABE}}\in\H_A\otimes\H_B\otimes\H_E$. Note that
this state has been obtained after interacting on half of a
maximally entangled state, so
\begin{equation}
\label{psiabe}
    \ket{\Psi_{ABE}}=(\one_A\otimes U_{BE})\,\frac{1}{\sqrt 2}
    (\ket{00}+\ket{11})\otimes\ket{E} .
\end{equation}
Eve's unitary operation acting on the qubit going to Bob and her
fixed ancillary system spans a two-dimensional subspace of
$\H_B\otimes\H_E$. It was shown in Ref. \cite{NG} that there exist
local bases $\ket{0'},\ket{1'}$ and $\ket{0},\ket{1}$ for Bob and
$\ket{0},\ket{1}$ for Eve such that
\begin{eqnarray}
\label{eveatt}
    U_{BE}\ket{0'}\ket{E}&=&\sin\alpha\ket{01}+\cos\alpha\ket{10}
    \nonumber\\
    U_{BE}\ket{1'}\ket{E}&=&\cos\beta\ket{00}+\sin\beta\ket{11} ,
\end{eqnarray}
where $0\leq\alpha,\beta\leq\pi/2$. Using the fact that $V\otimes
V^*\ket{\Phi^+}=\ket{\Phi^+},\,\forall\, V\in SU(2)$, we can take
on Bob's space the basis in the r.h.s. of Eq. (\ref{psiabe}) to be
the same as in the l.h.s. of Eq. (\ref{eveatt}). It follows that
all the states (\ref{psiabe}) can be easily parametrized as
\begin{equation}
\label{psiabef}
    \ket{\Psi_{ABE}}=\frac{1}{\sqrt 2}(\sin\alpha\ket{001}+\cos\alpha\ket{010}+
    \cos\beta\ket{100}+\sin\beta\ket{111}) ,
\end{equation}
i.e. they are completely specified by two angles, up to local unitary
transformations.

The state shared by Alice and Bob is
$\rho_{AB}=\Tr_E(\proj{\Psi_{ABE}})$. For any two-qubit state, the
maximal violation ${\cal B}$ of the CHSH inequality \cite{CHSH}
reads ${\cal B}=\sqrt{\lambda_1^2+\lambda_2^2}$, where the local
bound is put at one \cite{Horodecki}. The $\lambda_{1,2}$ are the
two largest (in modulus) eigenvalues of the $3\times 3$
correlation matrix $R(\rho)$ whose elements are
$(R(\rho))_{ij}=\Tr(\sigma_i\otimes\sigma_j\,\rho)$ where
$i,j=1,2,3$ and $\sigma_i$ denote the Pauli matrices. For the
state $\rho_{AB}$ one finds
\begin{equation}
\label{rab}
    R(\rho_{AB})=\left(\matrix{
    \cos(\alpha-\beta) & 0 & 0 \cr
    0 & \cos(\alpha+\beta) & 0 \cr
    0 & 0 & -\cos(\alpha+\beta)\cos(\alpha-\beta)}\right) .
\end{equation}
Note that $|R_{xx}|\geq |R_{yy}|\geq |R_{zz}|$, whence ${\cal
B}=(R_{xx}^2+R_{yy}^2)^{1/2}$ . After some simple algebra one
finds that
\begin{equation}
    {\cal B}>1\quad\Longleftrightarrow\quad 0\leq\alpha,\beta\leq\frac{\pi}{4},
    \mbox{  or  }\frac{\pi}{4}\leq\alpha,\beta\leq\frac{\pi}{2} .
\end{equation}

Now, let us see how the state $\rho_{AB}$ can be used for cryptography.
The honest partners measure in the local bases that are maximally correlated.
That is (see Eq. (\ref{rab})), Alice and Bob both measure in the $x$ basis.
Their measurement results are denoted by $\pm$, and the corresponding states
by $\ket{\pm}=(\ket{0}\pm\ket{1})\sqrt 2$. Since Alice's state is completely
random, $p_A(+)=p_A(-)=1/2$. Although Bob's state is different from $\one/2$,
we also have that $p_B(+)=p_B(-)=1/2$. Then $I(A:B)=I_b(R_{xx})$ where
$I_b$ is the binary mutual entropy
\begin{equation}
\label{iab}
    I_b(x)=1+\frac{1+x}{2}\log\left(\frac{1+x}{2}\right)+
    \frac{1-x}{2}\log\left(\frac{1-x}{2}\right) .
\end{equation}
Note that when $0\leq x_1,x_2\leq 1$, $I_b(x_1)\geq I_b(x_2)
\Leftrightarrow x_1\geq x_2$.

Eve's states, depending on Alice and Bob's results, read:
\begin{eqnarray}
    \ket{\tilde e_{++}}&=&\frac{1}{2\sqrt 2}\left((\cos\alpha+\cos\beta)\ket{0}+
    (\sin\alpha+\sin\beta)\ket{1}\right) \nonumber\\
    \ket{\tilde e_{+-}}&=&\frac{1}{2\sqrt 2}\left((-\cos\alpha+\cos\beta)\ket{0}+
    (\sin\alpha-\sin\beta)\ket{1}\right) \nonumber\\
    \ket{\tilde e_{-+}}&=&\frac{1}{2\sqrt 2}\left((\cos\alpha-\cos\beta)\ket{0}+
    (\sin\alpha-\sin\beta)\ket{1}\right) \nonumber\\
    \ket{\tilde e_{--}}&=&\frac{1}{2\sqrt 2}\left(-(\cos\alpha+\cos\beta)\ket{0}+
    (\sin\alpha+\sin\beta)\ket{1}\right) ,
\end{eqnarray}
the norm of the states being the probability of any event, i.e.
$p\,(00)=p\,(11)=(1+R_{xx})/4$ and $p\,(01)=p\,(10)=(1-R_{xx})/4$.
In the following, the tilde denotes non-normalized states. If Eve
wants to acquire information about Alice's result, she has to
distinguish between the two states $\rho_i=2(\proj{\tilde e_{i+}}+
\proj{\tilde e_{i-}})$, where $i=+,-$. In this case, the
measurement maximizing her information is known (actually, it also
minimizes her error probability) \cite{Peres}, having
$I(A:E)=I_b(\sin(\alpha+\beta))$. Therefore, $I_{AB}\geq I_{AE}$
when $R_{xx}=\cos(\alpha-\beta) \geq\sin(\alpha+\beta)$, and then
\begin{equation}
\label{infineq}
    I(A:B)>I(A:E)\quad\Longleftrightarrow\quad {\cal B}>1 .
\end{equation}
It is interesting to compute the information that Eve has about
Bob's symbol. Using the same techniques as above for the states
$\rho_i=2(\proj{\tilde e_{+i}}+ \proj{\tilde e_{-i}})$ where
$i=+,-$, one can see that $I(B:E)=
I_b(\sin\alpha\cos\alpha+\sin\beta\cos\beta)$. Note that
$I(A:B)>I(B:E)$ for all the values of $\alpha$ and $\beta$ but
$\beta=\pi/2-\alpha$, where the two quantities are equal
\cite{note0}. This means that the honest partners can apply a
reverse reconciliation protocol, i.e. one-way error correction and
privacy amplification from Bob to Alice, $\forall\,\alpha,\beta$,
except for a set of attacks of zero measure
($\beta=\pi/2-\alpha$). Eq. (\ref{infineq}) can now be extended to
\begin{equation}
\label{infineq2}
    I(A:B)>\max(I(A:E),I(B:E))\quad\Longleftrightarrow\quad {\cal B}>1
    .\label{res1}
\end{equation}
The entanglement properties of $\rho_{AB}$ also give more insight
into this result, since one can see that $\rho_{AB}$ is entangled
\cite{PPT} for all the attacks (except when $\beta=\pi/2-\alpha$).
Therefore none of the one-qubit attacks is able to disentangle Alice
and Bob.

\begin{center}
\begin{figure}
\epsfxsize=7cm \epsfbox{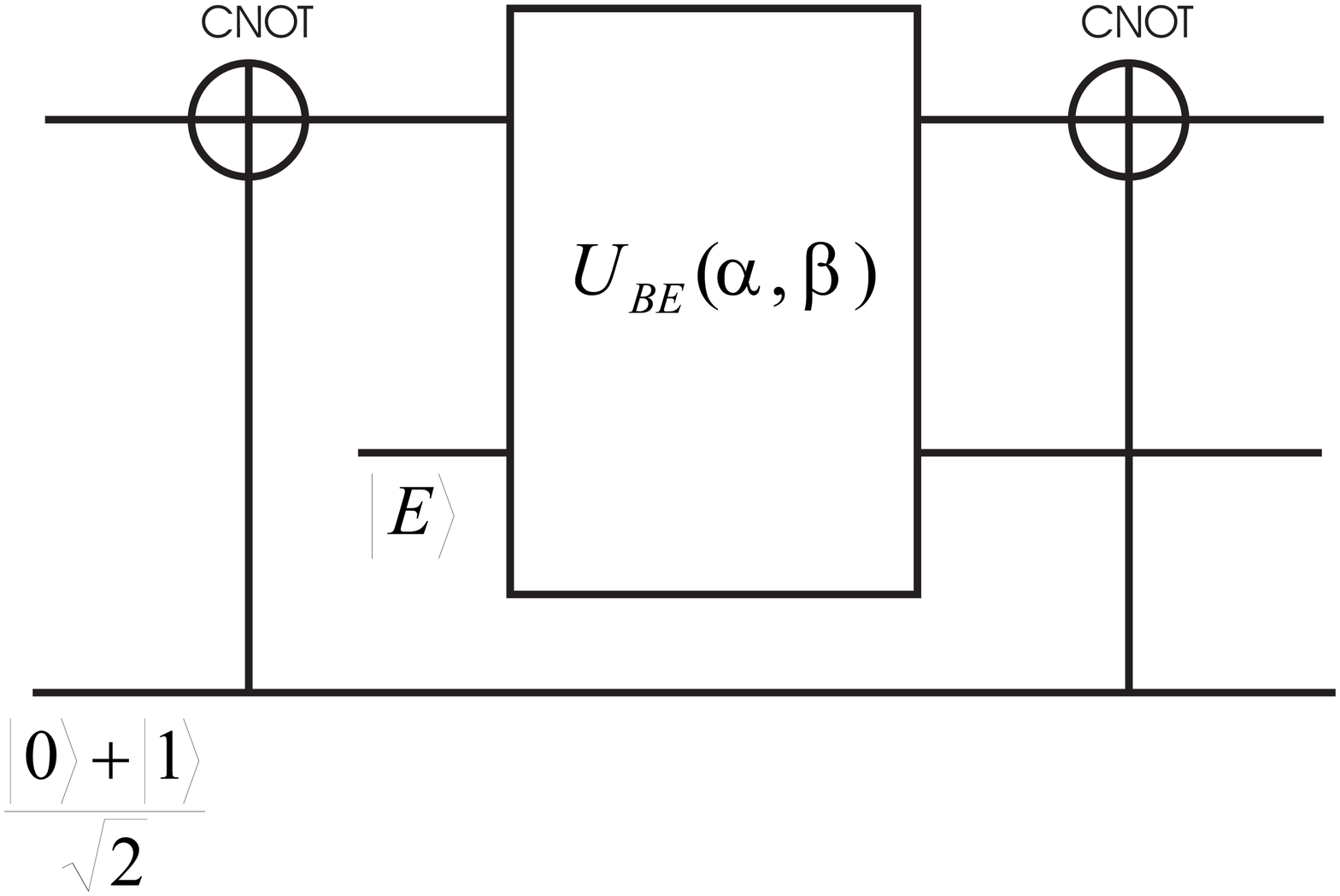} \caption{General scheme for
modifying the initial one-qubit attack, specified by
$U_{BE}(\alpha,\beta)$, where $I(A:B)\geq I(B:E)$ and $I(A:E)\geq
I(B:E)$, into a symmetric two-qubit attack, where $I(A:E)$ is the
same but now $I(B:E)=I(A:E)$.}\label{figcnot}
\end{figure}
\end{center}

There is a standard way in which Eve can make her information
about Alice and Bob symmetric, simply using the same one-qubit
attack, $U_{BE}(\alpha,\beta)$, and adding an extra ancillary
qubit. This symmetric two-qubit attack is shown in figure
\ref{figcnot}. The resulting state for Alice and Bob is
Bell-diagonal \cite{noteBell} and has the same correlations as
above, i.e. the same $R$ matrix (although $\rho_{AB}$ has now full
rank, while above its rank was equal to 2). Therefore, the
expression for the CHSH violation and the information Alice-Bob
has not changed. Concerning Eve, some simple and patient algebra
shows that her four two-qubit states are
\begin{eqnarray}
    \ket{\tilde e_{++}}&=&\frac{1}{2\sqrt 2}\left((\cos\alpha+\cos\beta)\ket{0}+
    (\sin\alpha+\sin\beta)\ket{1}\right)\otimes\ket{+} \nonumber\\
    \ket{\tilde e_{+-}}&=&\frac{1}{2\sqrt 2}\left((-\cos\alpha+\cos\beta)\ket{0}+
    (\sin\alpha-\sin\beta)\ket{1}\right)\otimes\ket{-} \nonumber\\
    \ket{\tilde e_{-+}}&=&\frac{1}{2\sqrt 2}\left((\cos\alpha-\cos\beta)\ket{0}+
    (\sin\alpha-\sin\beta)\ket{1}\right)\otimes\ket{-} \nonumber\\
    \ket{\tilde e_{--}}&=&\frac{1}{2\sqrt 2}\left(-(\cos\alpha+\cos\beta)\ket{0}+
    (\sin\alpha+\sin\beta)\ket{1}\right)\otimes\ket{+} ,
\end{eqnarray}
Now, it is easy to understand the role played by the second qubit. In the first
qubit we have the same information as above, so $I(A:E)$ has not changed.
From the second qubit Eve knows in a deterministic way whether Alice
and Bob symbol coincide. This allows her to use the knowledge on
Alice's symbol for guessing Bob's, and she now has $I(B:E)=I(A:E)$.
Thus, for this family of attacks
\begin{equation}
\label{infineq3}
    I(A:B)>\min(I(A:E),I(B:E))=I(A:E)=I(B:E)
    \quad\Longleftrightarrow\quad {\cal B}>1 .\label{res2}
\end{equation}
Since the present attack is symmetric, it is not important which
of the honest partners starts the one-way error correction and
privacy amplification processes.

In conclusion: for all individual attacks with just one qubit, the
link between sceurity and BI is given by (\ref{res1}). For the two
qubit attacks built from the one-qubit ones through the scheme of
Fig. \ref{figcnot}, the link is provided by (\ref{res2}). The
optimal individual eavesdropping on the BB84 protocol belongs to
this family of two-qubit attacks \cite{fuchs}. But we would like
to stress here that our results are independent of any considered
protocol. Indeed, we have studied the relation between Bell
violation and security for a family of states obtained after
eavesdropping on half of a maximally two-qubit entangled state. We
have shown that the violation of the CHSH guarantees the existence
of projective measurements whose results allow the honest partners
to establish a key with efficiency \cite{note1}. As expected,
these measurements are related to the bases that appear in the
violated CHSH inequality.

\section{Conclusions}

We have discussed the main connections between Bell's inequalities
and the usefulness of entanglement in quantum information
processing. In all the cases that have been considered, a state
that violates a Bell's inequality is useful for quantum
information processing; in the case of cryptography, it even leads
to efficient (one-way) secret-key extraction. Bell's inequalities
appear as detectors of ``efficient entanglement".

The precise link between entanglement, ``useful" or ``efficient"
entanglement, and non-locality remains however elusive, in spite
of all these clarifications. On the one hand, we are just now
beginning to tackle in a fruitful way the hard task of classifying
all the Bell's inequalities \cite{CGsliwa}. On the other hand, our
understanding of entanglement has been recently improved by a
remarkable result \cite{horo} by the Horodeckis and Oppenheim, who
have shown that a secret key can be extracted from some bound
entangled states: ultimately, we may discover that any form of
entanglement is ``useful" for something. This situation promises
still a lot of work to do for the future.


\section*{Acknowledgements}

We enjoyed several collaborations and discussions on these topics
with Daniel Collins, Michael Wolf, Marek \.{Z}ukowski and
\v{C}aslav Brukner. This work has been supported by the ESF, the
Swiss NCCR ``Quantum Photonics" and OFES within the EU project
RESQ (IST-2001-37559), the Spanish grant 2002FI-00373 UB and the
Generalitat de Catalunya.






\begin{thebibliography}{0}


\bibitem{EPR}
A. Einstein, B. Podolsky and N. Rosen, Phys. Rev. {\bf 47} 777
(1935).

\bibitem{Schr}
E. Schr\"odinger, Naturwissenschaften {\bf 23}, 807 (1935).

\bibitem{Bell}
J. S. Bell, Physics {\bf 1} 195 (1964).

\bibitem{AGR}
A. Aspect, P. Grangier and G. Roger, Phys. Rev. Lett.
{\bf 47}, 460 (1981).

\bibitem{loop} Both loopholes have been closed in separate
experiments. But at present, there has been no experiment
closing the two loopholes simultaneously.

\bibitem{QIPIntro98} H.K. Lo, S. Popescu, T.P. Spiller (eds), Introduction to Quantum computation and
information (World Scientific, 1998).

\bibitem{fuchs} C. Fuchs, N. Gisin, R.B. Griffiths, C.-S.
Niu, A. Peres, Phys. Rev. A {\bf 56}, 1163 (1997).

\bibitem{CK} I. Csisz\'ar, J. K\"{o}rner, IEEE Trans. Inf. Theory
{\bf IT-24}, 339 (1978).

\bibitem{Ekert} A. K. Ekert, Phys. Rev. Lett. {\bf 67}, 661 (1991).

\bibitem{qic} A. Ac\'{\i}n, N. Gisin, V.
Scarani, Quant. Inf. Comput. {\bf 3}, 563 (2003)

\bibitem{bell2} V. Scarani and N. Gisin, Phys. Rev. Lett. {\bf 87}, 117901 (2001);
idem, Phys. Rev. A {\bf 65}, 012311 (2002).

\bibitem{main} A. Ac\'{\i}n, V. Scarani, M.M. Wolf, Phys. Rev. A {\bf 66},
042323 (2002); idem, J. Phys. A: Math. Gen. {\bf 36}, L21 (2003).

\bibitem{wwzb} R. F. Werner and M. M. Wolf, Phys. Rev. A {\bf 64} 032112
(2001); M. \.{Z}ukowski and \v{C}. Brukner, Phys. Rev. Lett. {\bf
88} 210401 (2002).

\bibitem{Uffink}
J. Uffink, Phys. Rev. Lett. {\bf 88}, 230406 (2002).

\bibitem{complex} \v{C}. Brukner, M. \.{Z}ukowski, A. Zeilinger, Phys. Rev. Lett. {\bf
89}, 197901 (2002); \v{C}. Brukner, M. \.{Z}ukowski, J.-W. Pan, A.
Zeilinger, quant-ph/0210114.

\bibitem{NG} C.-S. Niu and R. B. Griffiths, Phys. Rev. A {\bf 60}, 2764 (1999).


\bibitem{CHSH} J. F. Clauser, M. A. Horne, A. Shimony, R. A.
Holt, Phys. Rev. Lett. {\bf 23} 880 (1969).

\bibitem{Horodecki} R. Horodecki, P. Horodecki and M. Horodecki, Phys. Lett. A
{\bf 200}, 340 (1995).

\bibitem{Peres} A. Peres, {\em Quantum Theory: Concepts and Methods} (Kluwer,
Dordrecht, 1998).

\bibitem{note0} This point was missed in the introductory
paragraphs of Refs. \cite{bell2}. The mistake is found in eq. (5)
of the Phys. Rev. A paper: $D_{BE}$ should read $\demi(1-\demi\sin
2\phi)$. Fig.1 of both papers suffer of this mistake: actually,
for the one-qubit attack studied there,
$\min(I_{AE},I_{BE})=I_{BE}$, which is in turn always smaller than
$I_{AB}$. The main results of these papers, concerning
multi-partners cryptography (where it is not even clear how
``reverse reconciliation" should be defined) are not invalidated;
nor is the message of the introduction, since, as we are going to
show, the symmetry $I_{AE}=I_{BE}$ is recovered by giving Eve a
second qubit.

\bibitem{PPT} Since $\rho_{AB}$ is a two-qubit state, its entanglement can be
detected by the non-positivity of the partial transposition,
$\rho_{AB}^{T_A}$. The partial transposition, that was introduced
in Ref. \cite{parttr}, is defined as follows: consider an
operator, $X$, that acts on $\H_A\otimes\H_B$, where $d_A$ and
$d_B$ denote the dimension of each space. The partial trasposition
of $X$ with respect to the first subsystem, in the basis
$\{\ket{1},\ldots,\ket{d_A}\}$, is given by $X^{T_A}\equiv
\sum_{i,j=1}^{d_A}\sum_{k,l=1}^{d_B}\bra{ik}O\ket{jl}\ket{jk}\bra{il}$.

\bibitem{parttr} A. Peres, Phys. Rev. Lett. {\bf 77} 1413 (1996); M. Horodecki, P.
Horodecki and R. Horodecki, Phys. Lett. A {\bf 223} 1 (1996).

\bibitem{noteBell} A state is said to be Bell diagonal when its eigenbasis is the
Bell basis, i.e. the four two-qubit maximally entangled states
$\ket{\Phi^\pm}=(\ket{00}\pm\ket{11})/\sqrt 2$ and
$\ket{\Psi^\pm}=(\ket{01}\pm\ket{10})/\sqrt 2$.

\bibitem{note1} Alice and Bob obtain the secret key by measuring always in the $x$
basis. However, Eve cannot pass unnoticed by using an
intecept-resend strategy, because under such an attack the CHSH
inequality ceases to be violated. In practice, Alice and Bob
should (seldom but randomly) modify their measurement, in order to
check the violation of CHSH in a subsequent sifting procedure. If
the quantum channel allowed the violation, then they can extract a
secure key; otherwise, they abort the protocol.

\bibitem{CGsliwa} C. S{\l}iwa, quant-ph/0305190;
D. Collins and N. Gisin, quant-ph/0306129.

\bibitem{horo} K. Horodecki, M. Horodecki, P. Horodecki and J. Oppenheim,
quant-ph/0309110.


\end{thebibliography}
\end{document}